\documentclass[draft,grl]{agutex}

 \usepackage{lineno}
  \usepackage[dvips]{graphicx}
\usepackage{rotating}
  \usepackage{multirow}
  \usepackage{url}
  \setkeys{Gin}{draft=false}
\authorrunninghead{NINA ET AL.}

\titlerunninghead{DETECTION OF IONOSPHERE RESPONSE ON GRB\lowercase{S}}

\authoraddr{Corresponding author: A. Nina,
Institute of Physics, University of Belgrade, Pregrevica 118, 11080 Belgrade, Serbia.
(sandrast@ipb.ac.rs)}

\begin{document}

%
%

\title{Detection of short term response of the low ionosphere on Gamma Ray Bursts}
%
%

%
%



 \authors{Aleksandra Nina,\altaffilmark{1}
 Sa\v{s}a Simi\'{c},\altaffilmark{2} Vladimir A. Sre\'ckovi\'c,\altaffilmark{1} and
 Luka \v{C}. Popovi\'{c}\altaffilmark{3,4}}

\altaffiltext{1}{Institute of Physics, University of Belgrade, Pregrevica 118, 11080 Belgrade, Serbia.}

\altaffiltext{2}{Faculty of Science, Department of Physics, University of Kragujevac, Radoja Domanovi\'{c}a 12, 34000 Kragujevac, Serbia.}

\altaffiltext{3}{Astronomical Observatory, Volgina 7, 11060 Belgrade, Serbia.}

\altaffiltext{4}{Faculty of Mathematics, Department for Astronomy, University of Belgrade, Studentski trg 16, 11000 Belgrade, Serbia}

%
%


\begin{abstract}
In this paper, we study the possibility of detection of short term terrestrial
lower ionospheric response to gamma ray bursts (GRBs) using a statistical analysis of perturbations
of six very low or low frequency (VLF/LF) radio signals emitted by transmitters located worldwide and recorded by VLF/LF receiver located in Belgrade
(Serbia). We consider a sample of 54 short lasting GRBs (shorter than 1 min) detected by the SWIFT satellite during the period 2009-2012. We find that a statistically significant perturbations can be present in the low ionosphere, and reactions on GRBs may be observed immediately after the beginning of the GRB event or with a time delay of 60 s - 90 s.
\end{abstract}

%
%

%

\begin{article}

%
%

\section{Introduction}
Gamma-Ray Bursts (GRBs) are known as the most energetic phenomena in the Universe where a huge amount of energy is released
and their investigation is of a great astrophysical importance
(see for review \cite{geh09} and references therein).
Consequently, there is a question: how much can a GRB event disturb the
Earth atmosphere, especially the ionosphere? It is expected that photons with a minimum energy of a few keV
can influence the electrical conductivity of the ionosphere, however the very low ionization cross section for these photons in the ionosphere may provide an effective transit of GRBs without significant perturbation in the ionosphere.

On the other hand the recording of
a GRB event in the ionosphere is a very complex task since many physical phenomenon perturb the ionosphere including lightning, electron precipitation, solar activity, and more. Also, GRBs and observed part of ionosphere may have different characteristics during a GRB event. Therefore, it is very
hard to prove that a particular perturbation, observed close to a GRB  event, is related to it.

The ionospheric perturbation caused by a particular GRB can be confirmed if it is sufficiently intensive and long-lasting
as reported by \cite{ina07a} that the atmosphere exposed to the impact of radiation was disturbed more than one hour at altitudes below about 70 km. The modeling of electron density perturbations for this case shows that the variation of this parameter increases toward the lower heights and reaches rise of 4 orders of magnitude at an altitude of about 20 km.
Although satellites have recorded an average of about 1 GRB per day, these intensive ionospheric reactions are very rare and only a few such events have been recorded. An indication of a long term ionospheric perturbation was first reported by \cite{fis88}
 for the GRB30801 event.
They proposed that a network of VLF signal
monitors may provide some relevant information on GRBs and their influence on the ionosphere.
However, after this report, there are only a few papers which reported the detection of long term ionospheric perturbations
due to GRB events: GRB030329 \citep[][]{mae05},
  GRB041227 \citep{ina07a,hua08}, GRB 060124 \citep{hud10}.  GRB090122 \citep{tan10},  GRB080320A \citep{hud10} and GRB080319D
  \citep{slo11}.  An effort for the indirect detection of GRBs by their ionospheric response  observed by VLF/LF signals,
 and discussion on its possible impact on the GRB science and investigations in general, has been given in \cite{slo11}.

However, there are no investigations of low intensive ionospheric perturbations. The aim of this paper is to present possible short duration ionospheric perturbations caused by GRB events. We used VLF (3 kHz - 30 kHz) and LF (30 kHz - 300 kHz) radio signals to detect ionosphere disturbances in a short period around GRB events for 54 GRBs observed by SWIFT during 2009 - 2012.

\section{Observations}
\label{motivation}

 In this study, we are trying to explore any perturbation in the ionosphere which may be connected with a GRB event and we are using a statistical analysis to find ionospheric perturbations in the period of a GRB event. Consequently, we search for any ionospheric perturbation around the GRB event.
To determine time intervals around a GRB event recording by satellite we first analyzed several studies related to the low ionospheric reactions to GRBs.

 \subsection{The Observed Time Intervals and Duration of Ionospheric Perturbations}
 During X-ray flares, there is a time delay between the start of satellite recording and the moment when perturbation is observed in the ionosphere. One can expect the same to be true for GRBs. Moreover, this was confirmed during the
  GRB041227 event \citep[see][]{ina07a}, and  GRB080319D and GRB080320A events \citep{hud10}.

  The durations of ionospheric perturbations, presented in pervious works, connected with GRB events are usually long term as can be seen, for example, in \cite{ina07a}  (longer than one hour) and \cite{cha10}  (several seconds) and they depend on considered location \citep{tan10}. However, short term perturbations are also presented during relevant long term reaction \citep{ina07a}. Here we consider that short duration ionospheric perturbations last less than 1 s and others we consider as long term. Therefore, for our investigation we use the VLF/LF data with a time resolution of 0.02 s.

\begin{figure}
\noindent \includegraphics[width=0.95\columnwidth,angle=0]{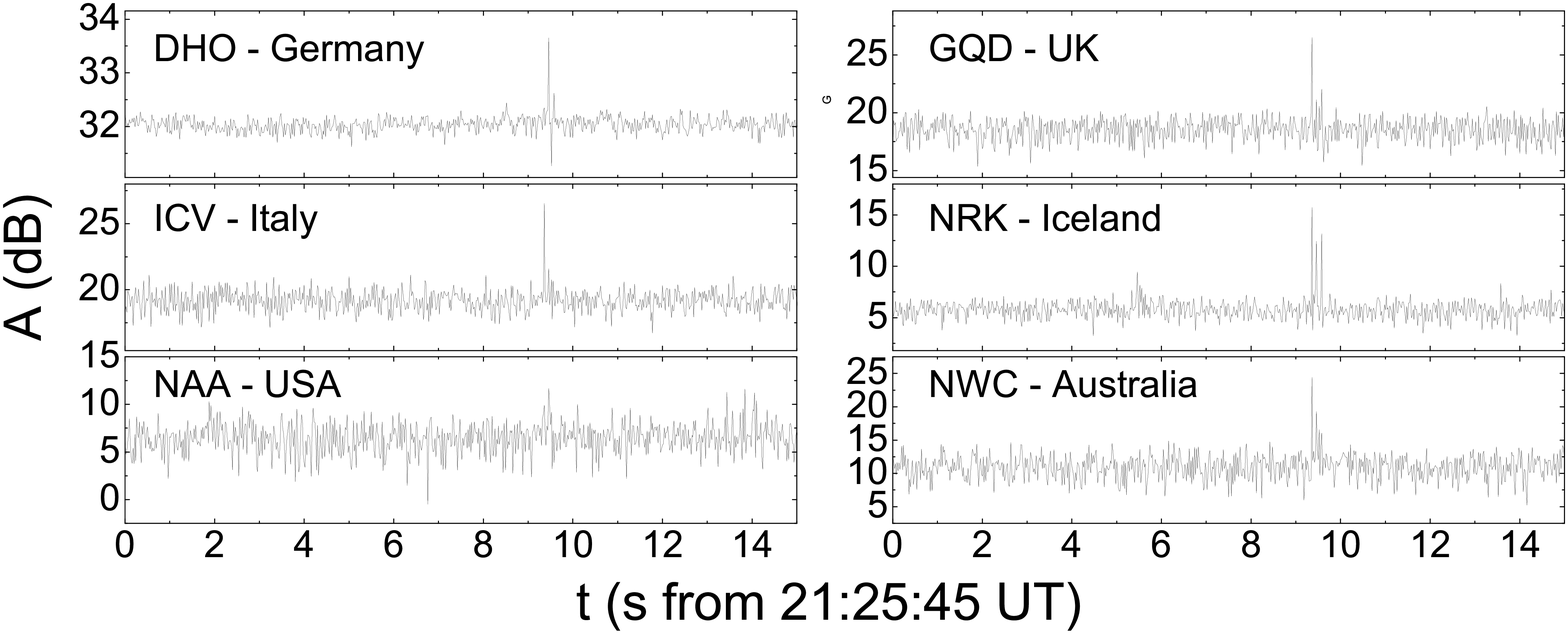} \\ \includegraphics[width=0.75\columnwidth,angle=0]{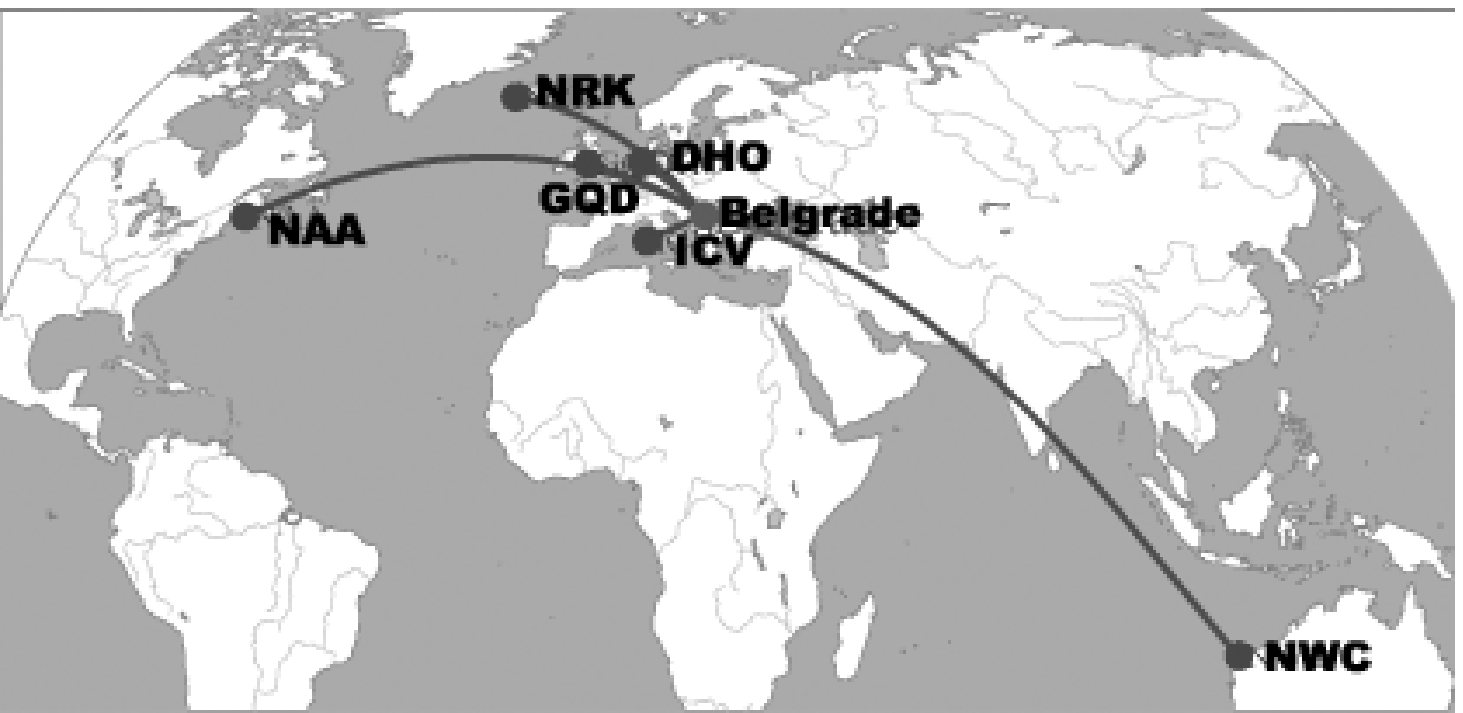}
\caption{Upper panel: A short-time amplitude amplifications in the VLF/LF signals emitted from six transmitters (labeled at panels) and recorded by the Belgrade VLF/LF receiver possibly induced by the GRB110223B event occurred at 21:25:48 UT on Feb 23, 2010 less than 10 s after its beginning. Bottom panel: The transmitters and receiver locations and signal propagation paths are indicated in the map. Transmitters and signal characteristics are given in Table \ref{tableR1}.}\label{MIusekundu}
\end{figure}

One should consider that the ionospheric perturbations can appear during and after a GRB event, we consider an interval of 4 min around the GRB event, i.e. 2 min before and 2 min after the start of the GRB event recorded by the SWIFT satellite. It allows us to compare perturbations that occur immediately before and after the beginning of he GRB event.
 Fig. \ref{MIusekundu}  shows an example of short term  amplitude perturbations of signal (upper panels) emitted by transmitters shown in the map (bottom panel) and recorded with Belgrade receiver less than 10 s after the GRB110223B event observed at 21:25:48 UT on Feb 23, 2010.

\subsection{Experimental Setup}

A low ionospheric electron density disturbance perturbs subionospheric VLF/LF signals which propagate
in the earth-ionosphere waveguide on Great Circle Paths
(GCPs).  The reflection of VLF/LF signal occurs below 85 km and depends on the local electron density. The electron density indicates ionospheric perturbation:
this causes time variations of the VLF/LF wave
trajectory and, consequently, recorded wave amplitude
and phase \citep{gru08}.

The amplitude and phase of
VLF/LF radio signals observed at any point can thus be used to measure the spatial
and temporal characteristics of local disturbances in the lower ionosphere.
The advantage of this method is that it provides continuous monitoring of a large portion of the lower ionosphere. This is achieved with VLF/LF receivers monitoring the signal, with $<$ 1 second time resolution, from VLF/LF transmitting beacons which are scattered around the world. These signal perturbations allow recording periodic and global influences as well as unforeseen and located events  (for more details see Rfs \cite{ina90,tho00,hal06,nin13a,sal13}).

In this paper we used six radio signals in VLF and LF domains emitted from DHO (Germany), GQD (UK),
ICV (Italy), NRK (Iceland), NAA (USA) and NWC (Australia)
 and recorded by the VLF/LF receiver located in Belgrade (Serbia) (see map in Fig. \ref{MIusekundu}). The
 characteristics of transmitters, signal propagation paths and related amplitudes
are given in Table \ref{tableR1}.

We utilize the VLF/LF amplitudes of these transmitters recorded by the AWESOME (Atmospheric
Weather Electromagnetic System for Observation Modeling and
Education) receiver system \citep{coh10} located in Belgrade, Serbia (a part
of Stanford/AWESOME Collaboration for Global VLF Research)
with the sampling period of 0.02 s.

\section{Methods}
\label{metode}

\subsection{The GRB Sample}

In this study, we selected a GRB sample observed by the SWIFT satellite
when the received VLF/LF signal is not significantly affected by other perturbers.
This satellite contains the
Gamma-ray telescope which covers energy range from 15 to 150 keV, X-ray telescope monitoring from 0.2 to
10 keV and Ultraviolet/optical instrument for afterglow detection. In this study we consider the first phase of GRB events, the so-called gamma
phase, with photon energies in GeV and MeV down to keV band and a time interval
less then 100 s (for the most of bursts). We used data obtained by the Gamma-ray telescope related to the parts of hard X (12.4 keV - 124 keV) and $\gamma$ radiation (above 124 keV). The time resolution of these data is 0.01 s.
 To analyze the whole duration of the GRBs and possible secondary reactions after the intrusion of intense  the high-energy radiation we selected only short lasting GRBs, i.e. GRB events lasting less than 1 min. From the obtained data the maximum lasting of a GRB event in the sample is around 50 s.

To estimate the GRBs duration in the selected sample we used the so-called radiation intensity criteria, i.e. we start measurements at the moment when the radiation of pulses in a GRB lightcurve exceeds background noise by 10\%.
The duration is then measured in such a way to include all of the existed peaks in the observed lightcurve. This approach is good
enough for our purpose, since we expect that sudden or sharp increase of the radiation could possibly produce some ionization events. Using the criteria mentioned above, we are able to select 54 GRBs in the period 2009-2012 (see Table S1).

\subsection{Signal Processing}

\begin{figure}
 \noindent\includegraphics[width=0.55\columnwidth,angle=0]{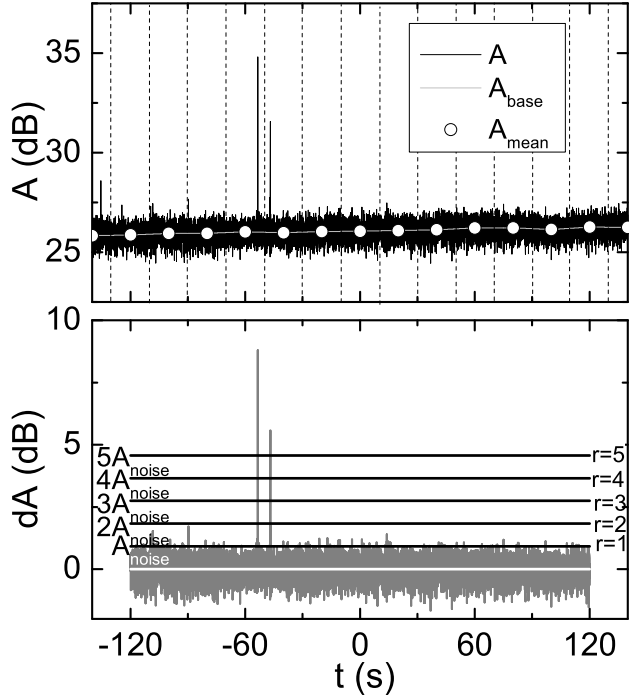}
\caption{The method for peak determination in a VLF/LF signal. Upper panel: Determination of baseline amplitude, $A_{\rm {base}}$, (gray line) by interpolation of averaged amplitude, $A_{\rm {mean}}$, (scatters) within 20 s intervals. Bottom panel: Determination of levels $r$ that is 2 to 5 times larger then noise level. Amplitude noise $A_{\rm {noise}}$ is defined as the maximum absolute value of $dA$ (deviation of recorded amplitude $A$ of baseline amplitude $A_{\rm {basea}}$) after elimination $p=2\%$ of its largest and lowest values. }\label{sumovi}
\end{figure}

In this study we developed a method for accounting peaks in the VLF/LF signal. We compare the signal 2 min before and 2 min after a GRB to find a number of peaks before and after the GRB event.
The tendency of signal amplitude and its noise depends on signal characteristics and properties of medium where it is propagating. Therefore, first we found unperturbed signal amplitude, the so-called baseline, $A_{\rm {base}}$ as it is illustrated  in Fig. \ref{sumovi}, upper panel. To find the noise level we calculated deviations of recorded amplitude $A(t)$ from $A_{\rm {base}}$ ($dA(t)=A(t)-A_{\rm {base}(t)}$). Amplitude noise $A_{\rm {noise}}$ is defined as the maximum absolute value of $dA$ after elimination $p=2\%$ of maximum and minimum values of $dA$ (see Fig. \ref{sumovi}, bottom panel). The detailed analysis shows that the noise level dependent on $p$ affects only the value $r$ (see Eq. \ref{eq:1prvo}) at which the signal response detections start without essential effects on visualization of detections (see Fig. S2 and its explanation in Supporting Information). In the present analysis we fix $p$ to 2\% to provide better visibility without loss of significant information.

Peak classification is determined quantitatively by the ratio of deviations of the recorded amplitude $A(t)$ from the amplitude of the base curve $A_{\rm {base}}(t)$ at the time $t$, and the noise amplitude $A_{\rm {noise}}$ as:
  \begin{linenomath*}
  \begin{equation}
    \label{eq:1prvo}
    {
    \frac{A(t)-A_{\rm {base}}(t)}{A_{\rm {noise}}}\geq r
    }
  \end{equation}
  \end{linenomath*}
 for $r=$ 2, 3, 4 and 5 (see Fig. 2 bottom panel). In the analysis we found that there is no big difference between number of peaks for $r=3$ and 4 and 5, therefore here we will present the results for $r= 2$ and 3.

To find statistical significance of appearing peaks in time-interval (shown in Figs \ref{histogrami1} and \ref{histogrami2}) we calculated

    \begin{linenomath*}
    \begin{equation}
      \label{eq:1drugo}
      {
      \sigma_i=\frac{x_{\rm {i}}-\overline{X}}{\overline{X}}\cdot100\%,
      }
    \end{equation}
    \end{linenomath*}
    where $x_{\rm {i}}$ is corresponding number of peaks in the bin $i$ and $\overline{X}$ is an averaged number of peaks in whole interval, i.e.:

     \begin{linenomath*}
     \begin{equation}
      \label{eq:1trece}
      {
      \overline{X}=\frac{1}{N}{\sum_i}x_i,\quad i=1,...,N.
      }
     \end{equation}
     \end{linenomath*}
In addition, only values of $\sigma_{\rm {i}}$  higher than 15\% in the case of 30 s bins and 100\% higher in the case of 5 s bins are marked with black circles in Figs \ref{histogrami1} and \ref{histogrami2}.

It is well known that ionosphere is different during daytime and nighttime conditions. Therefore, our sample is additionally divided into three subsamples: signals obtained during daytime and nightime and periods when the solar terminator (ST) affects considered ionosphere part resulting into a non-stationary behavior of the recorded VLF/LF signal amplitude (see Figs \ref{histogrami1} and \ref{histogrami2}).

Additionally we apply technique as given in \cite{ina07b}. However, contrary to the case of \cite{ina07b}, the analysis of our sample shows that this method cannot account short term and temporary impacts from ionospheric natural fluctuations (see Supporting Information Fig. S1). Therefore we will consider our method in further analysis.

  \section{Results and Conclusions}

Here we explore the possibility of GRB influence on the low ionosphere by forming a sample of 54 GRBs and accounting the number of peaks in signal amplitudes before and after beginning of GRB events recorded by SWIFT satellite. We analyze signals from six transmitters (see Table \ref{tableR1}) Figs 3 and 4.

\begin{figure}
\noindent\includegraphics[width=0.45\columnwidth,angle=0]{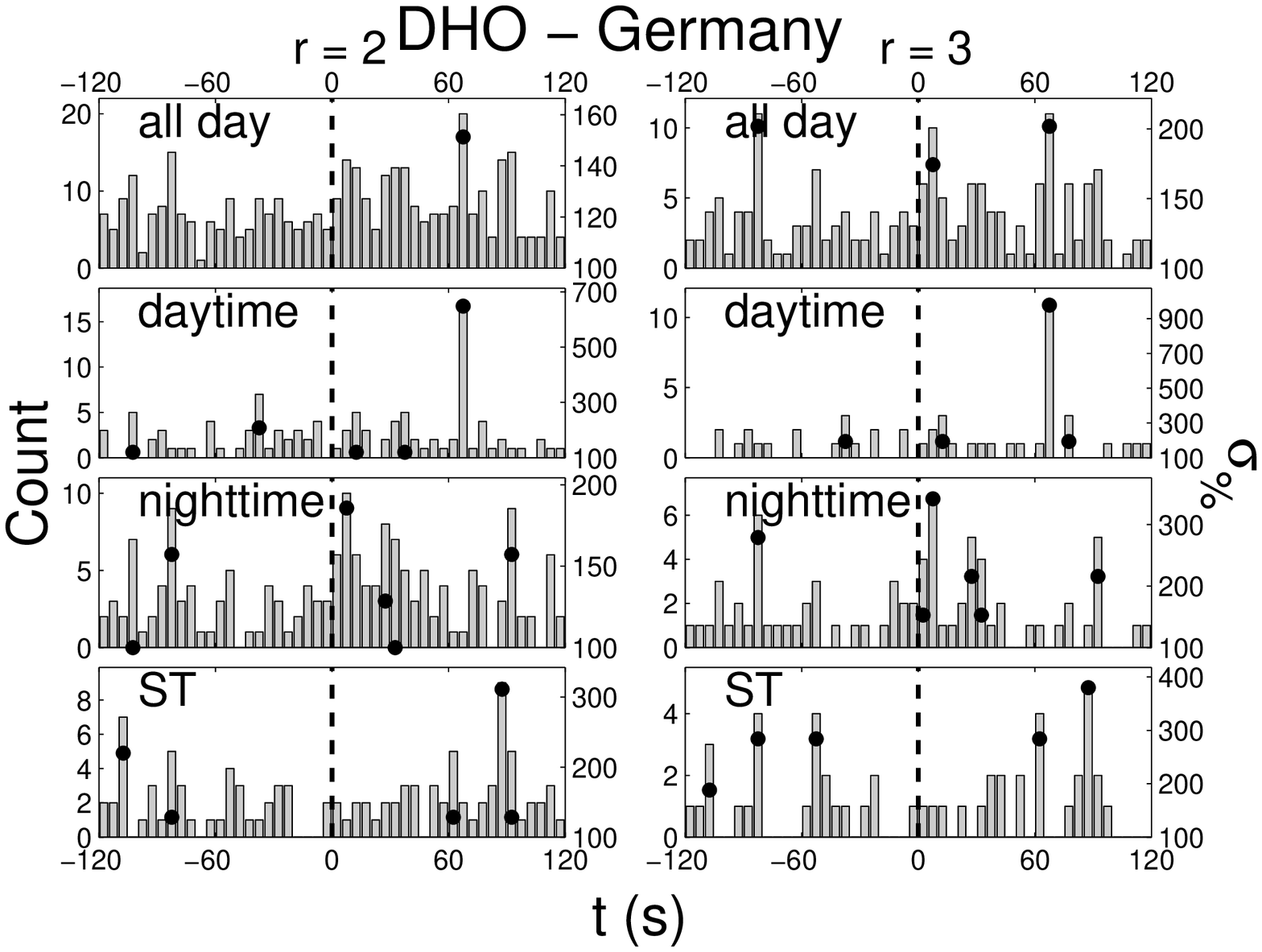}\includegraphics[width=0.45\columnwidth,angle=0]{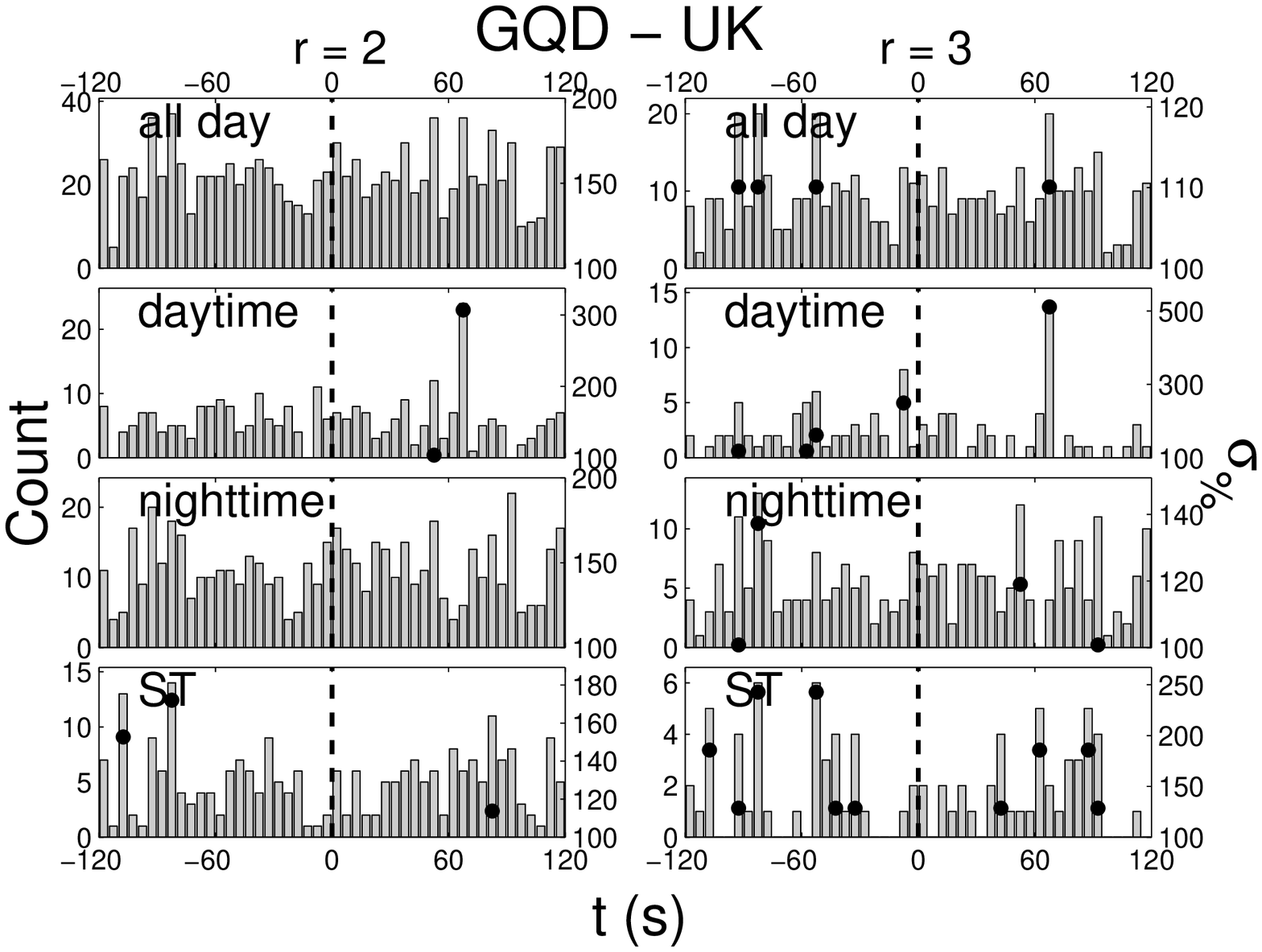}\\
\noindent\includegraphics[width=0.45\columnwidth,angle=0]{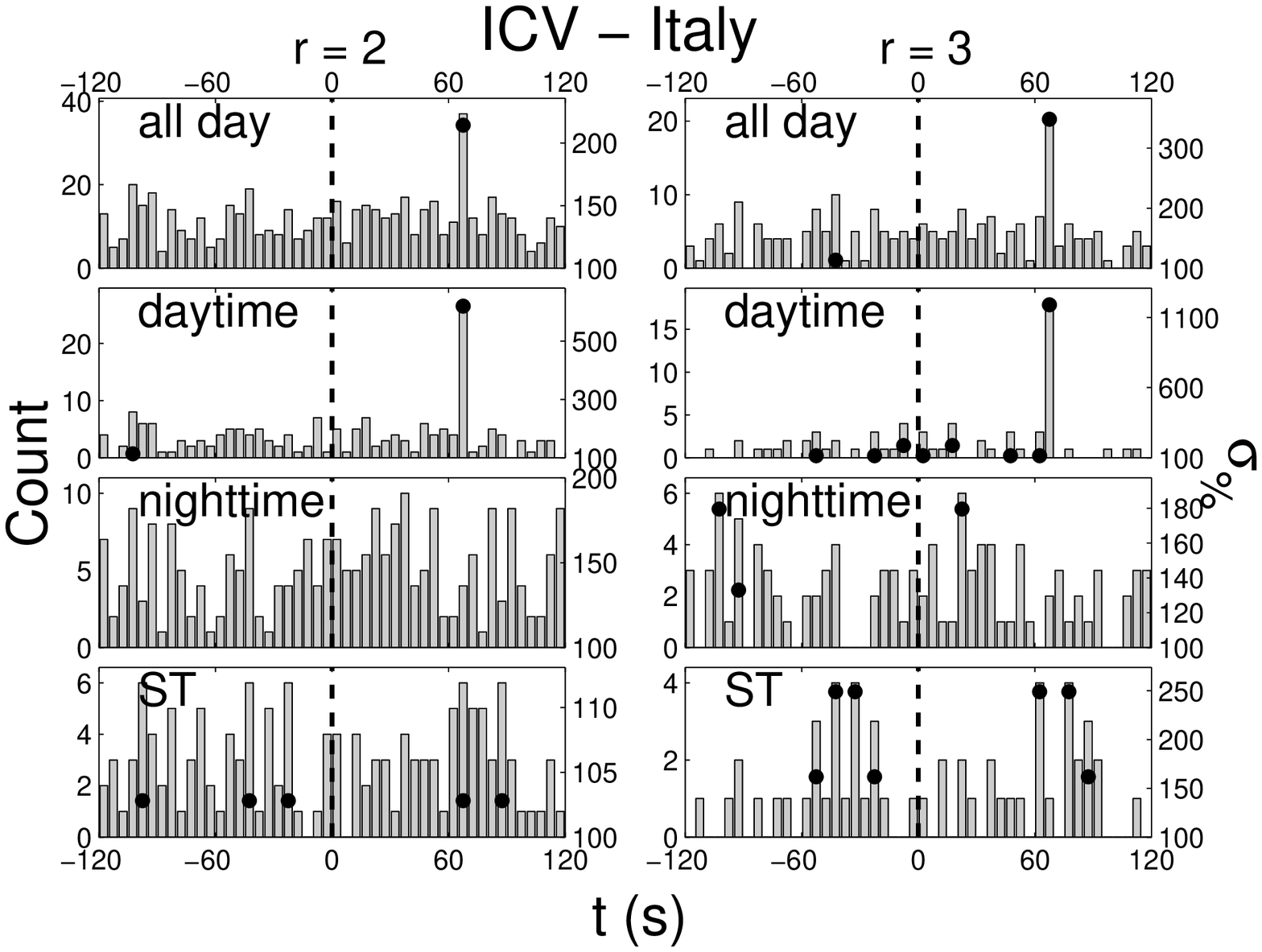}\includegraphics[width=0.45\columnwidth,angle=0]{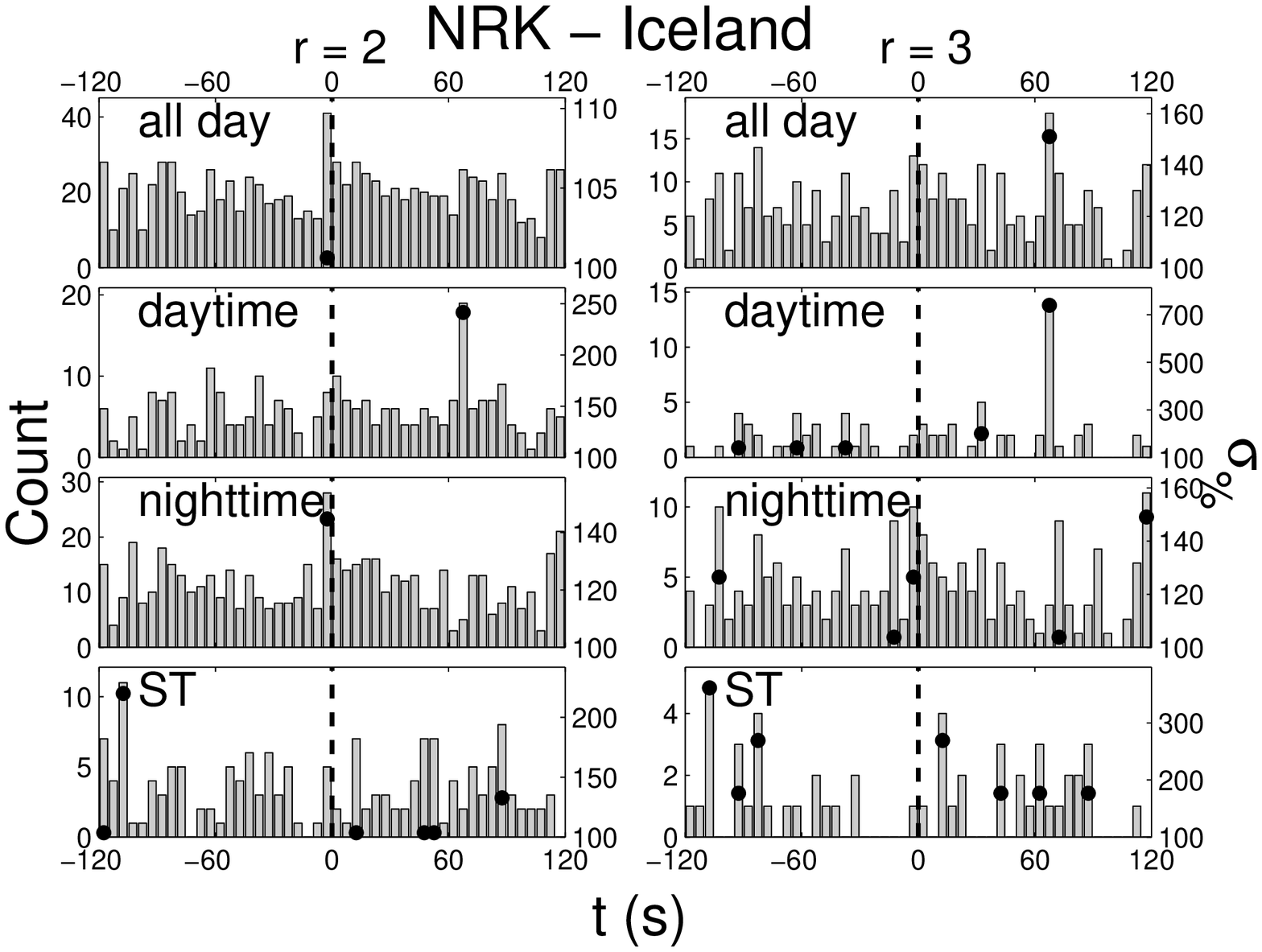}\\
\noindent\includegraphics[width=0.45\columnwidth,angle=0]{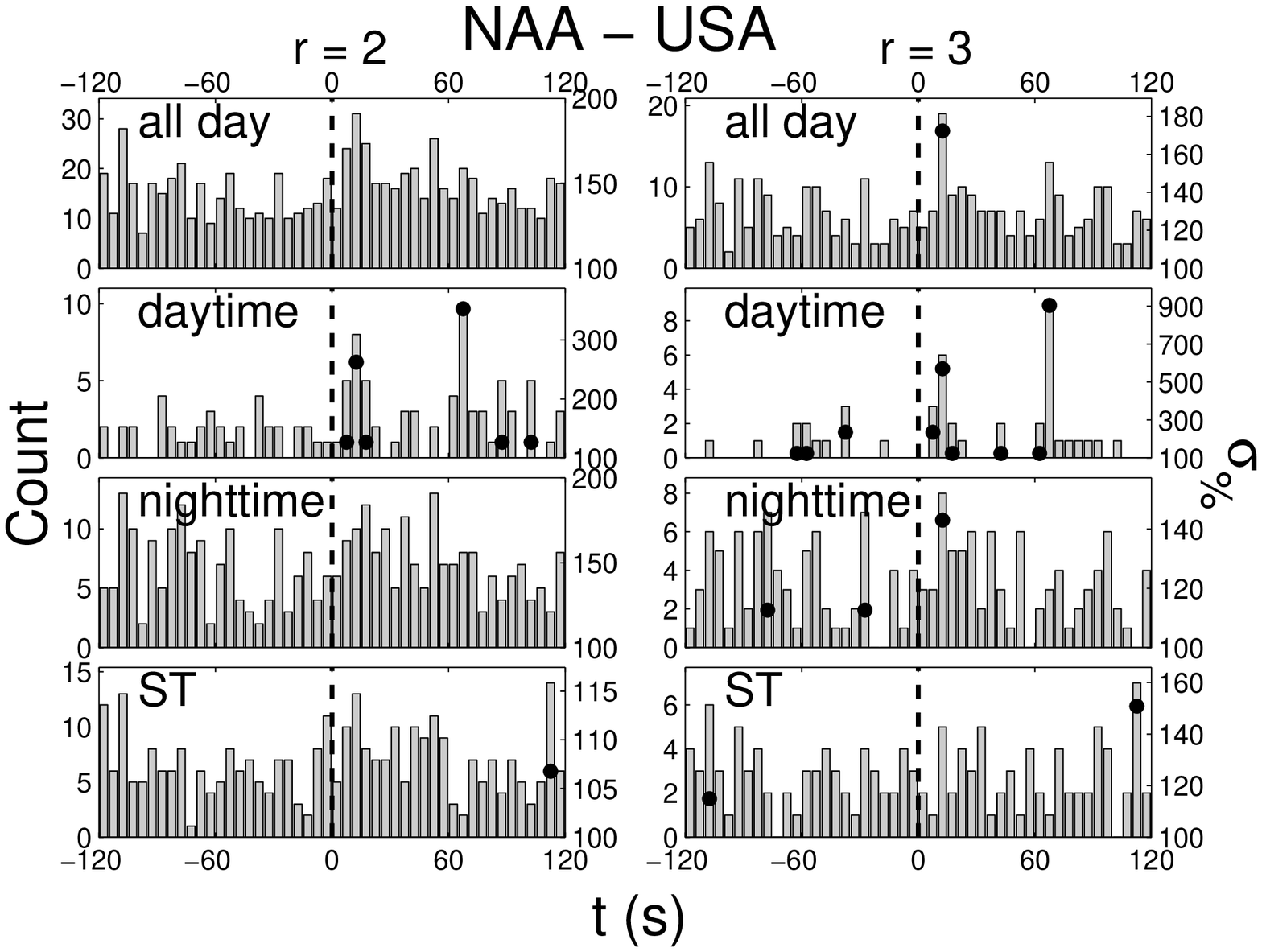}\includegraphics[width=0.45\columnwidth,angle=0]{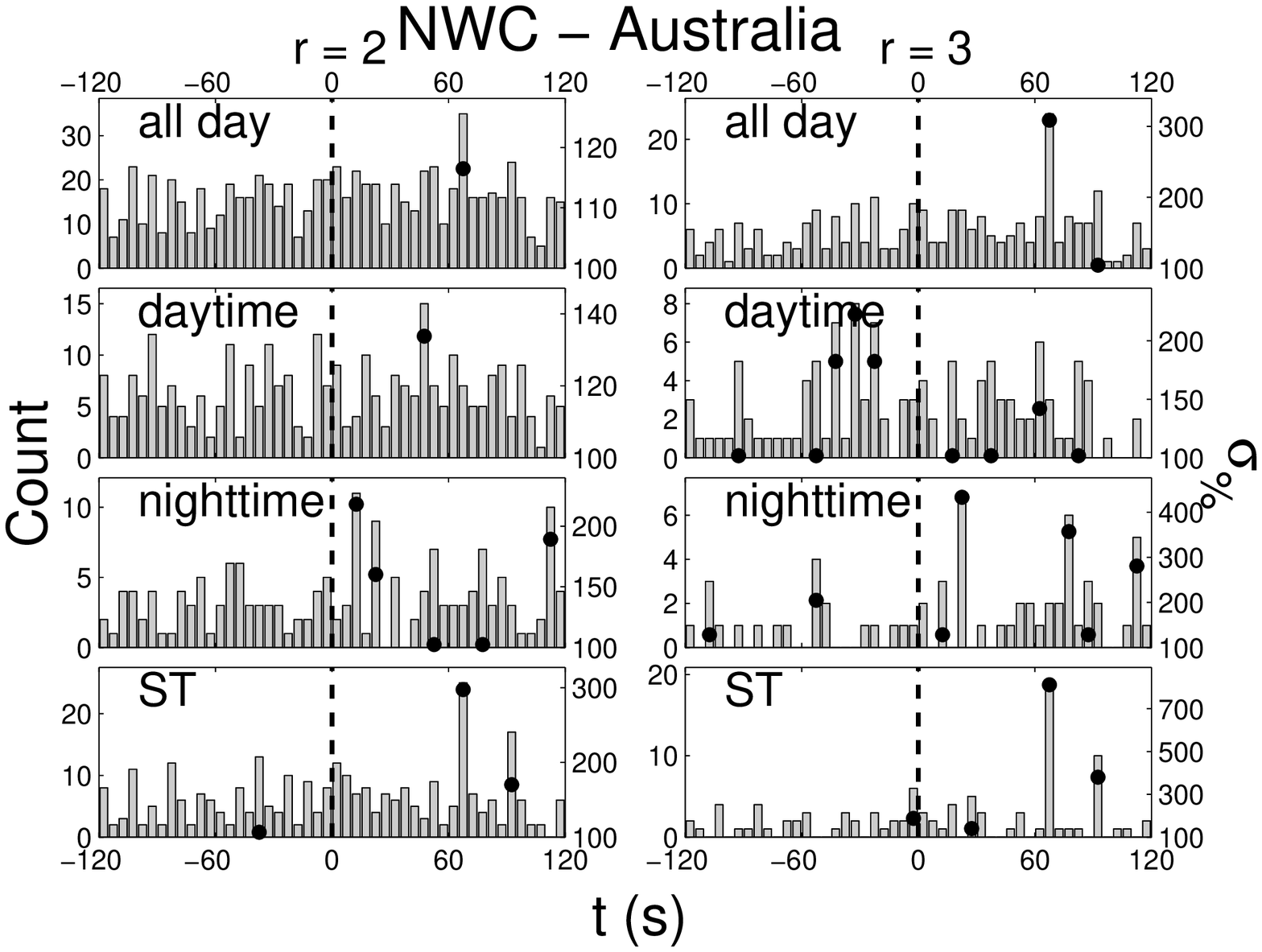}
\caption{Histograms of number of peaks for signals emitted by different transmitters (noted on plots) accounted in time bins of 5 s for $r=2$ (left panels) and $r=3$ (right panels). The vertical lines denote the time of the satellite GRB recording and the black points indicate the statistically significant increase of peaks numbers within bins indicated by values of $\sigma$ given by Eq. \ref{eq:1drugo} (right axes). For each transmitter the whole sample, daytime, nighttime and ST conditions are given (from top to bottom).}
 \label{histogrami1}
\end{figure}

\begin{figure}
\noindent\includegraphics[width=0.45\columnwidth,angle=0]{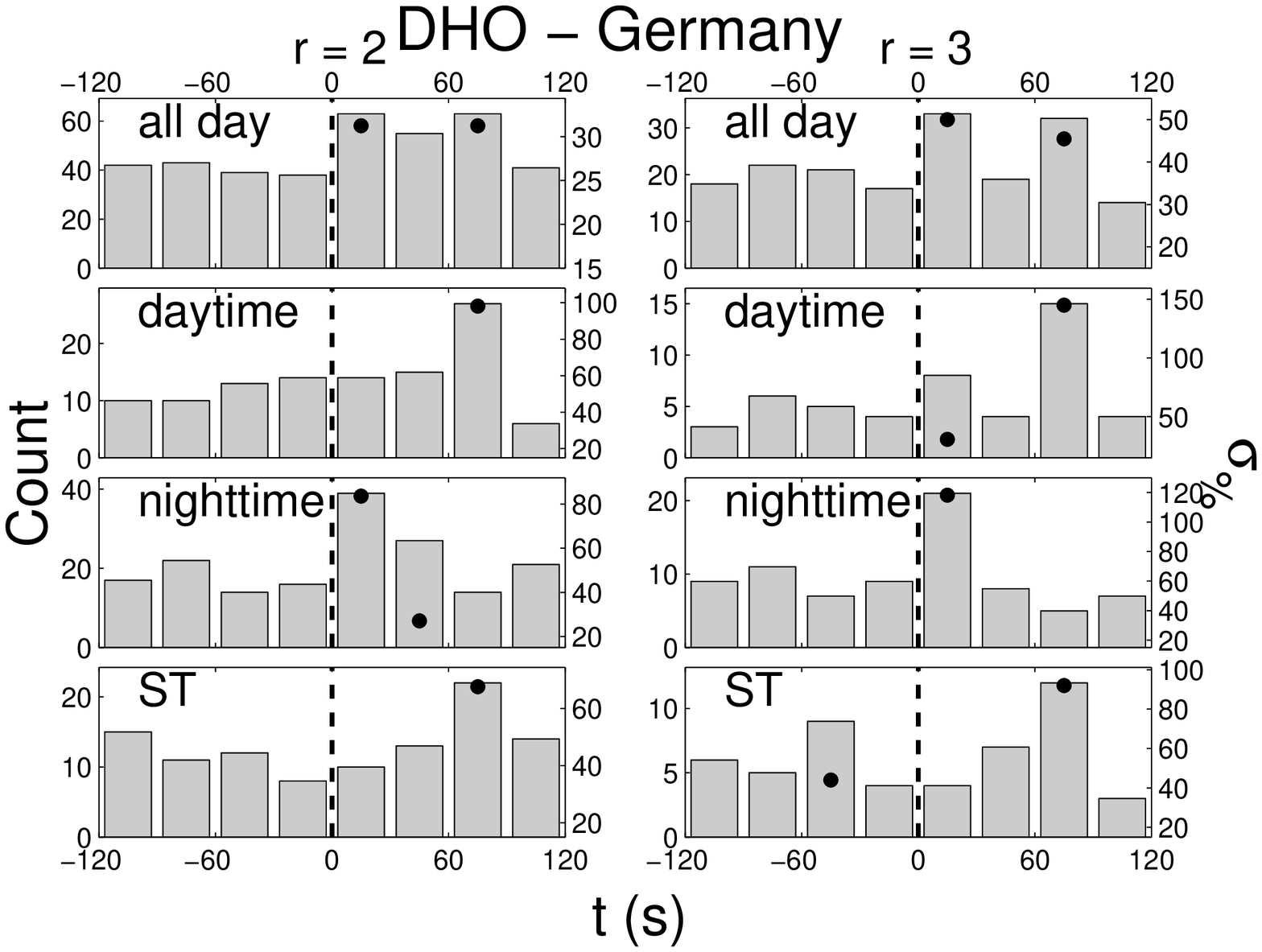}\includegraphics[width=0.45\columnwidth,angle=0]{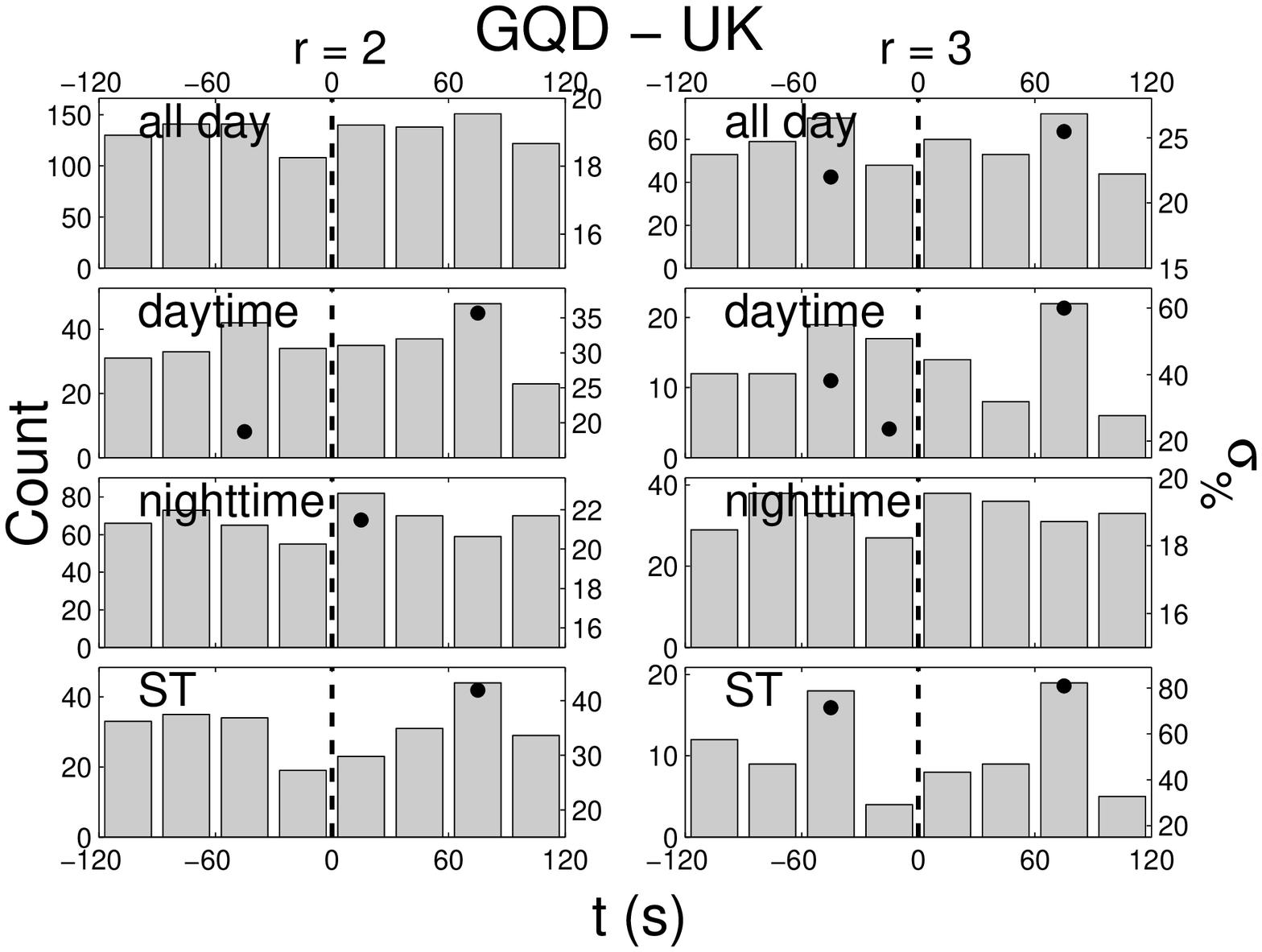}\\
\noindent\includegraphics[width=0.45\columnwidth,angle=0]{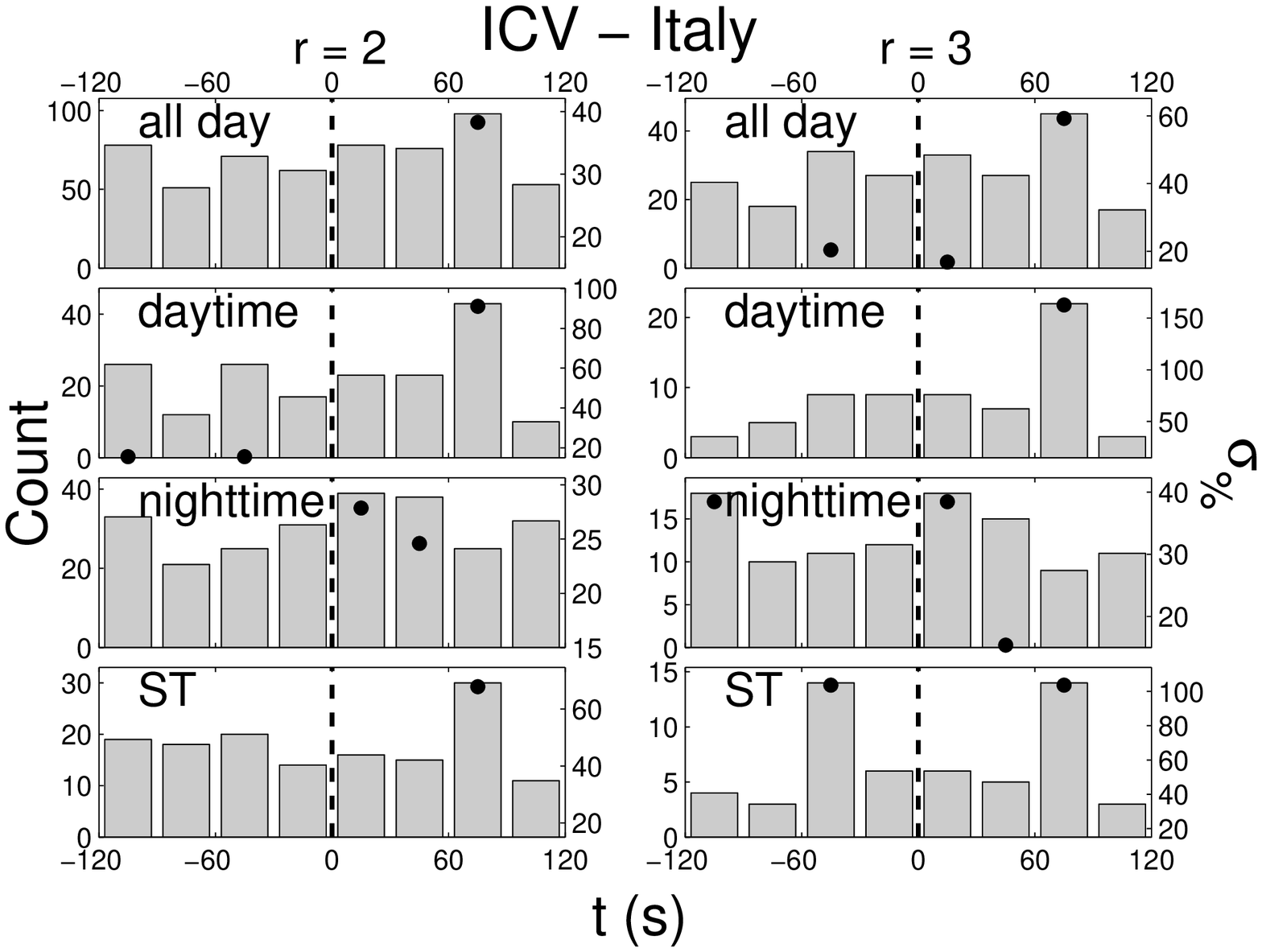}\includegraphics[width=0.45\columnwidth,angle=0]{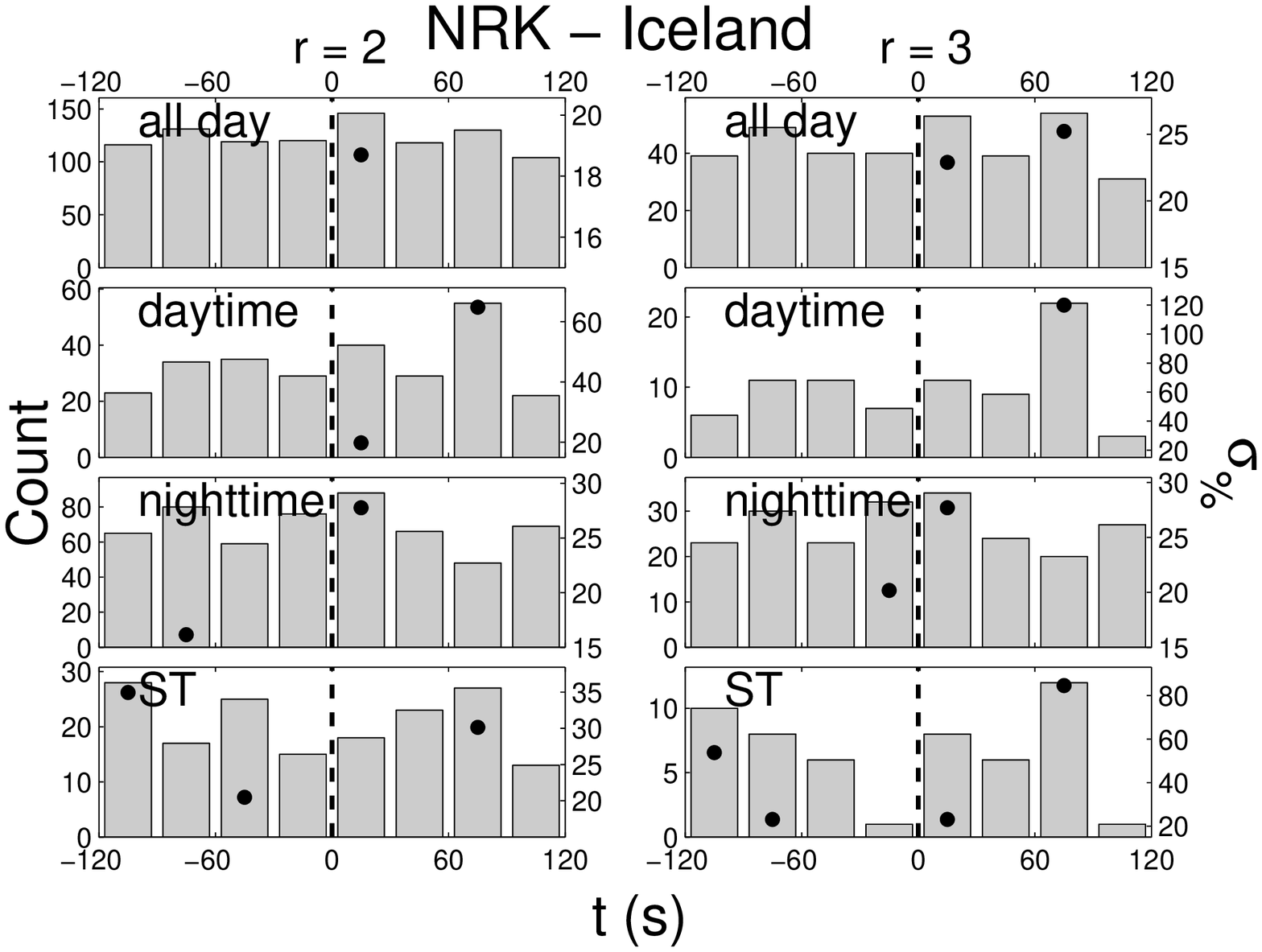}\\
\noindent\includegraphics[width=0.45\columnwidth,angle=0]{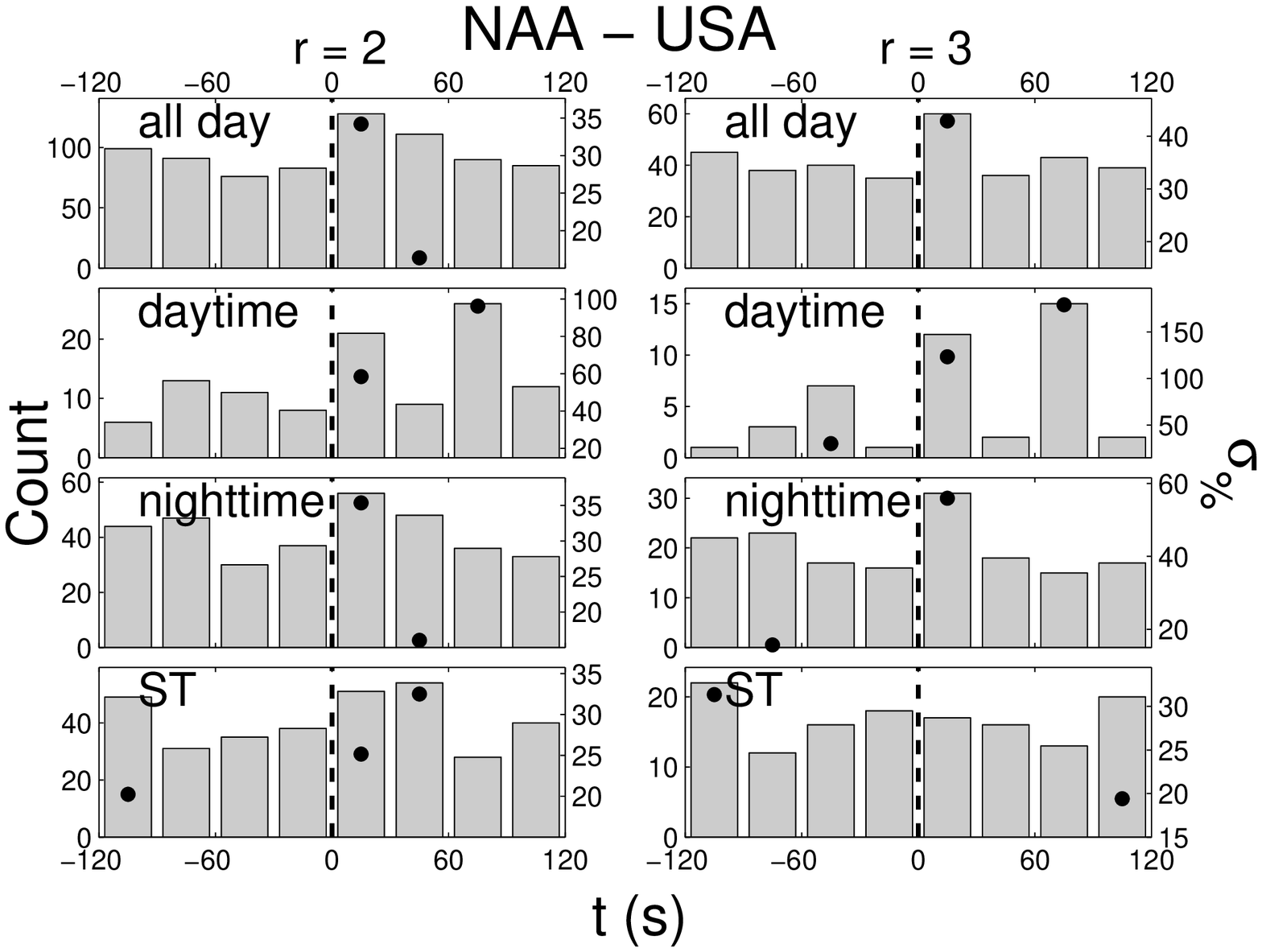}\includegraphics[width=0.45\columnwidth,angle=0]{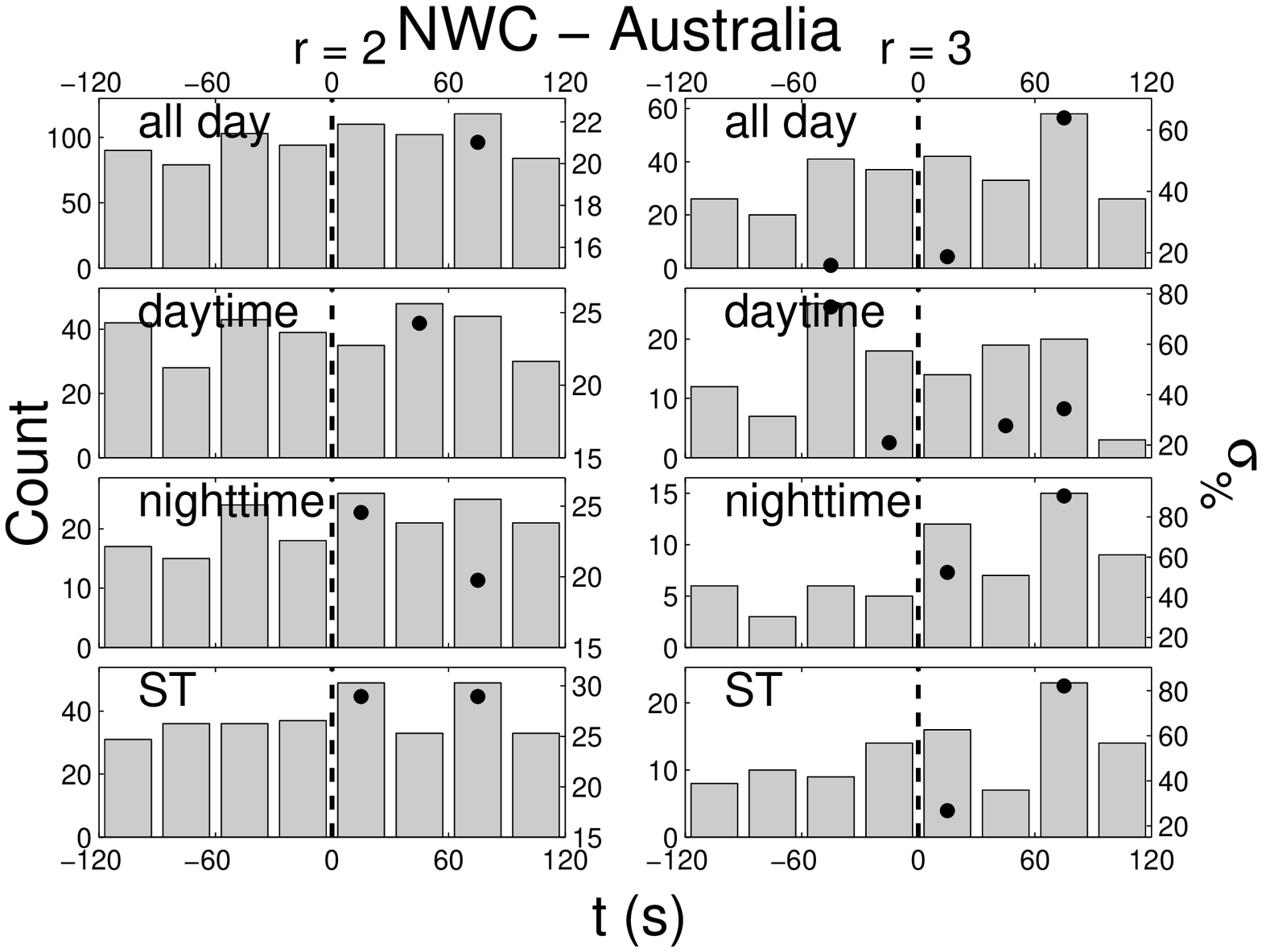}
\caption{The same as in Fig. \ref{histogrami1} but for wider bins of 30 s.}
 \label{histogrami2}
\end{figure}

As can be seen in Figs \ref{histogrami1} and \ref{histogrami2} we show histograms of number of recorded amplitude peaks within the interval of 120 s before and after the GRB detections for bins time interval of 5 s and 30 s, respectively. Additionally, we separately considered daytime, nighttime and ST subsamples (see Figs \ref{histogrami1} and \ref{histogrami2}). The statistical analysis is given for the considered 6 signals recorded in Belgrade and 54 GRB events (except for the DHO signal from Germany that was used for 51 cases). Here we consider the number of peaks with $r=$ 2 and 3 times larger amplitude than the noise amplitude $A_{\rm {noise}}$ (for particular transmitter panel left and right, respectively). For a particular transmitter panels from top to bottom we show the whole sample, daytime and ST conditions, respectively.

As can be seen in Fig. \ref{histogrami1} in the case of narrow bins (NB), two time domains (TD) with jumps in the histograms can be noticed: 1. immediately after the beginning of satellite recording of GRB (NB-TD1) and 2. some time after the beginning of the GRBs (NB-TD2).

\begin{enumerate}
\item \textbf{NB-TD1.} In the period after the satellite recordings of GRB, the peaks are noticeable for signals from Germany and Australia (nighttime conditions), and USA (both daytime and nighttime conditions).

\item \textbf{NB-TD2.} Increase of amplitude peak numbers after intrusion of the observed radiation, recorded within interval 65 s - 70 s, is noticeable in all signals.

In any case, the analysis for NB-TD1 and NB-TD2 shows that the presence of considered reaction is statistically important in histograms with bin width of 5 s and that there exists a certain time interval from the recording of GRBs by satellite until the detected ionospheric perturbations.
\end{enumerate}

The increase in number of amplitude peaks during and after GRB events is clearly seen in all six transmitter signals when the bin width is expanded from 5 s to 30 s (wide bins - WB). In this case one can notice (see Fig. \ref{histogrami2}) the following:
\begin{enumerate}
\item \textbf{WB-TD1.} In the first thirty seconds, the most important jumps in the number of amplitude peaks are again recorded in all cases within the day period that depends on the signal. Also, the most important increases of intensive amplitude peaks are recorded during nighttime (for signals from Germany, Italy and USA) in periods 30 s - 60 s after a GRB beginnings.
\item \textbf{WB-TD2.} Increase of amplitude peaks number within the period of 60 s - 90 s from the beginning of the GRBs recording by satellite is significant for all transmitters in the daytime condition. Also, it is present during periods of ST influence on all signals except those from USA (for this signal increase is present in the next bin), and Australia during nighttime.
\end{enumerate}
This definitely confirms the presence of observed short term perturbations and their appearance after the start of the GRB detection by satellite.

Comparison of results for different bin widths shows more clear confirmation of the GRB detectability by VLF/LF radio signals for the wider temporal bins which suggests that intensification of the ionospheric plasma ionization starts with different "delays" with respect to the satellite GRB recording. This conclusion and the occurrence of relevant peaks in the histograms after GRB duration indicate possible secondary processes that affect ionization in the low ionosphere.

Finally, we can point out the most important conclusion of this study: it confirms detectability of a short term reaction of the low ionosphere to GRBs which does not cause intense long term reactions in general. The important perturbations of the low ionospheric plasma occur at different times in relation to the beginning of GRB events which indicates a possibility of detection of ionization by the primary GRB radiation as well as by some of its secondary effects. The presented study indicates the possibility of intensive ionospheric perturbations during the whole day.

Obtained results in this paper show that a detailed analysis of GRB influence on the ionosphere is needed. This study will be in focus of our upcoming research. In addition, we want to point out that presented model for extraction of intensive peaks from the noise and further processing can be applied to different events.


%
%
%
%
%
%
%

\begin{acknowledgments}
The data for this paper collected by SWIFT satellite
are available at NASA's Goddard Space Flight Center.  Data set name: Old Swift Trigger and Burst Information (for 2009-2012).     \url{http://gcn.gsfc.nasa.gov/swift2009_grbs.html}, \url{http://gcn.gsfc.nasa.gov/swift2010_grbs.html},
\url{http://gcn.gsfc.nasa.gov/swift2011_grbs.html},
\url{http://gcn.gsfc.nasa.gov/swift2012_grbs.html}.
The used data are also given in Supporting Information Table 1. Requests
for the VLF/LF data used for analysis can be directed to the
corresponding author.

The authors are thankful to the Ministry of Education, Science and Technological Development
of the Republic of Serbia for the support of this work within the projects 176001, 176002 and III44002.

The authors
are also grateful to Vladimir M. \v{C}ade\v{z}, Morris Cohen and an anonymous referee for very useful suggestions and comments.
\end{acknowledgments}

\end{article}

 \begin{sidewaystable}
\caption{Characteristics of transmitters, signal propagation paths and number of GRB events related to VLF/LF signals propagation during daytime, nighttime and ST conditions.}
\begin{tabular}{l l l l|l|l l l}
\hline
\multicolumn{4}{c|}{\textbf{TRANSMITTER}} & \multicolumn{1}{c|}{\textbf{SIGNAL PATH}} & \multicolumn{3}{c}{\textbf{NUMBER OF GRBs}}\\
\hline
\multicolumn{1}{c}{Sign} & \multicolumn{1}{c}{Location} & \multicolumn{1}{c}{Frequency (kHz)} & \multicolumn{1}{c|}{Power (kW)} & \multicolumn{1}{c|}{Length (km)} & \multicolumn{1}{c}{daytime} & \multicolumn{1}{c}{nighttime} & \multicolumn{1}{c}{ST} \\
\hline

DHO &  Rhauderfehn, Germany & 23.4 & 800 & 1304 & 13  &  28   & 10 \\
GQD & Anthorn, UK & 22.1 & 200 & 1935  & 16 &  28   & 10 \\
ICV & Isola di Tavolara, Italy & 20.27 & 20 & 976 & 16   & 28  & 10 \\
NRK & Grindavik, Island & 37.5 & 800 & 3230 & 16  &  28  &  10 \\
NAA & Cutler, Maine, USA & 24.0 & 1000 & 6548 & 11  &  20  &  23 \\
NWC & North West Cape, Australia & 19.8   & 1000 & 11974 & 15  &  8  &  31 \\
\hline
\end{tabular}
\label{tableR1}
\end{sidewaystable}

\end{document}